\documentclass[sn-mathphys-ay]{sn-jnl}

\usepackage{graphicx}%
\usepackage{multirow}%
\usepackage{amsmath,amssymb,amsfonts}%
\usepackage{amsthm}%
\usepackage{mathrsfs}%
\usepackage[title]{appendix}%
\usepackage{xcolor}%
\usepackage{textcomp}%
\usepackage{manyfoot}%
\usepackage{booktabs}%
\usepackage{graphicx}
\usepackage{txfonts}
\usepackage{natbib}

\newcommand{\mbar}{\overline{M}}
\newcommand{\rbar}{\overline{r}}

\begin{document}
\title[The RR Lyrae distribution in the Galactic Bulge]{The RR Lyrae distribution in the Galactic Bulge}


\author*[]{\fnm{Roberto} \sur{Capuzzo-Dolcetta}}\email{roberto.capuzzodolcetta@uniroma1.it}



\affil[]{\orgdiv{Department of Physics}, \orgname{Sapienza, università di Roma}, \orgaddress{\street{2 Piazzale A. Moro}, \city{Roma}, \postcode{00185},  \country{Italy}}}






\abstract{\textbf{Purpose:} RR Lyrae stars are important distance indicators. They are usually present in globular clusters where they were first discovered. The study of their properties and distribution in our Galaxy and external galaxies constitutes a modern field of astrophysical research. The aim of this paper is checking the possibility that the observed distribution of RR Lyrae stars in the Galactic bulge derives from orbitally decayed globular clusters (GCs).\\
\textbf{Methods:} To reach the aim of the paper I made use of the comparison of observational data of RR Lyrae in the Galactic bulge with the distribution of GCs in the Milky Way (MW) as coming from theoretical models under a set of assumptions.\\
\textbf{Results:} I obtain the expected numbers and distributions of RR Lyrae in the Galactic bulge as coming from an initial population of globular clusters at varying some characteristic parameters of the GC population and compare to observational data.\\
\textbf{Conclusion:} The abundance of RR Lyrae distribution in the Galactic bulge and their radial distribution is likely still too uncertain to provide a straight comparison with theoretical models. Despite this, it can be stated that a significant fraction of the \lq foreground \rq ~RR Lyrae present in the MW originate from orbitally evolved and dissolved GCs.}




\maketitle

\section{Introduction}
\label{sec:1}
RR Lyrae stars are variable stars commonly found in globular clusters (GCs), constituting a set of variables distinct from the classical Cepheids. Actually, RR Lyrae are shorter in periods, poorer in metallicity and differently spatially located (in the Galaxy) respect to Cepheids. They are spread over all latitudes, as expected because they are Pop. II stars, contrary to Cepheids which are mostly confined to the Galactic disk.
Even if they are often referred to as "cluster variables", their presence is not limited to GCs, even if a significant fraction of the presently observed RR Lyrae in the Milky Way (MW) actually belongs to GCs. This high fraction raises the question of what is the origin of the RR Lyrae observed out of clusters, which is the main topic of this paper.
Their average luminosity is lower than classical Cepheid's, meaning that their use as distance indicators cannot be extended to very large distances in spite of the tight period-luminosity correlation, at least in the infrared K-band \citep{cat15}.  Recently, \citet{chen23} demonstrated that double-mode RR Lyrae can be used as distance indicators in the near-field universe at accuracy better than $3 \%$. Several other works, such as \cite{coh17}, \cite{her18} and \cite{ior19} also used RR Lyrae to study the MW stellar distribution in the disc and in the halo up to 140 pc.

Regarding the number of RR Lyrae in the Galactic bulge and their kinematic properties we cite \citet{Sos14}, \citet{Sos19}, \citet{Alc98}, \citet{Du20} and \citet{Kun22}. 

Due to their faintness, a reliable determination of the distribution of type ab RR Lyrae (RRab, hereafter also referred as RR or RR Lyrae) --by far the most common of the 3 types of RR Lyrae pulsators-- is only suitable for the Milky Way. The projected distribution of RRab stars from the Galactic  center to the halo has been obtained and discussed in \citet{Navetal21} by using a huge, $N_\textrm{RRab}=64,850$, sample from different surveys (VVV, OGLE and Gaia). One of the main results of this paper is that the RRab distribution favors the star cluster \textit{infall and merge} scenario for creating an important fraction of the central galactic region. The results of this paper have been used here to examine the intriguing possibility that a significant fraction of the RRab observed in the Galactic bulge is provided by GCs which have been carried to the inner Galactic region and there eventually partially dissolved.// The hypothesis of GC orbital decay in their motion in the parent galaxy has been investigated in several papers since \citet{trem75} with the aim also to explain the overdense central galactic regions \citep{Dolc93,amb} in a way alternative to the \textit{in situ} formation. In the framework of GC system evolution it looks as reasonable that part of their RR Lyrae content have been released to the Galactic environment and so the search of correlations among distribution of RR Lyrae in the bulge and that of GC can provide a proof of the validity of that model.
At this regard, it is relevant citing \citet{min23} who presented a discussion of the RR Lyrae projected distribution from the Galactic center to the halo, providing relevant data for the scopes of this paper.

The paper is organized as follows: sect. 1 introduces to the topic; sect. 2 discusses the theoretical modelization and its methodological approach to the aim of the paper; sect. 3 presents and discusses the results. Finally, in sect. 4 there is a summary and a conclusive discussion.

\section{Method}
\label{sec:2}
In the assumption of initial uniform distribution of globular clusters over a spherical volume of radius $R$, and assuming a global content of $N_0$ RR Lyrae in the whole set of GCs, we can evaluate the number of RR Lyrae  belonging to GCs that, at the generic age $\tau$ of the sample, is contained within the sphere of generic radius $r$ around the galactic center as


\begin{equation}
\label{eq:1}
   N(<r;\tau) = N_{0} \left(\frac{r}{R}\right)^3\left(1+\frac{4\pi\int_{r}^R {\bar r}^2d\bar r\int\limits_{\mbar}^{M_M}\Phi(M) dM}{V(R)\int\limits_{M_m}^{M_M}\Phi(M)dM}\right),
\end{equation}

where $\Phi(M)$ is the RR Lyrae number abundance as function of the parent GC mass $M$, assumed non-zero in the $[M_m, M_M]$ interval and for $0\leq r \leq R$, so that $N_{0}= \int\limits_{M_m}^{M_M}\Phi(M) dM$ is the total number of RR Lyrae belonging to GCs of mass within the $[M_m,M_M]$ interval, $\mbar$ is the local dynamical friction (df) mass cut-off, and  $V(R)$ is the volume of the sphere of radius $R$ centered at the galactic center.

The dynamical friction cut-off mass $\mbar(\tau)$ is a (time dependent) threshold mass. Clusters moving on orbits of initial eccentricity $e$ in a Dehnen's $\gamma$ model \citep{Deh93} of a galaxy of mass and length scales $M_g$ and $r_g$, whose density distribution is

\begin{equation}
\label{eq:2}
\rho(r)=(3-\gamma)\dfrac{M_g}{4\pi r_g^3}\dfrac{1}{(r/r_c)^\gamma}\dfrac{1}{\left(r/r_g+1\right)^{4-\gamma}},
\end{equation}

and that are more massive than $\mbar$ are decayed by df within a given time $\tau$ from the Galactocentric distance $\bar r$ to $r$ while lighter clusters are still on the way.

\begin{figure}
	\includegraphics[scale=0.6]{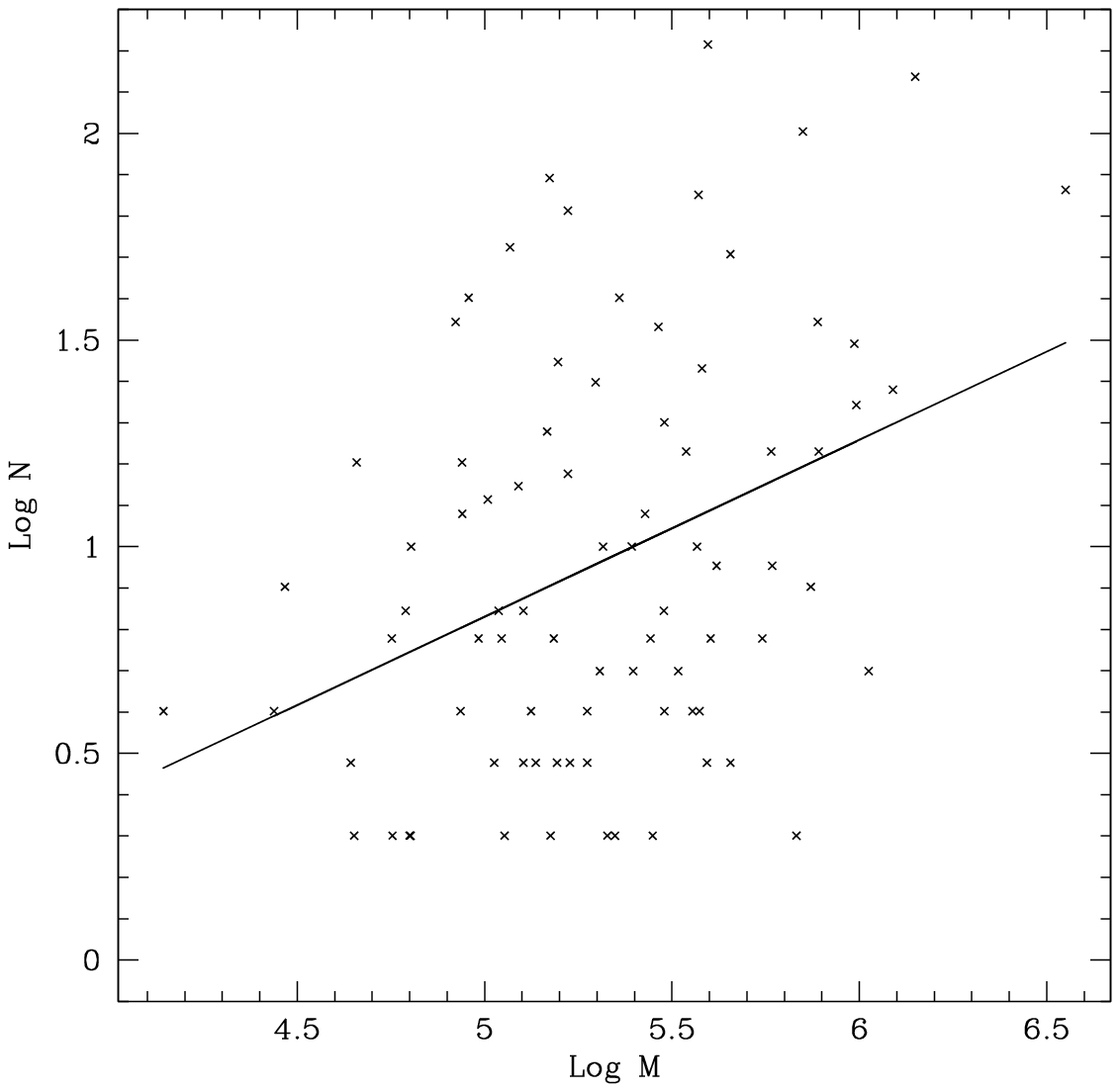}
    \caption{The number of RRab contained in Galactic globulars as function of the hosting GC mass (in solar masses). Data for RRab are from \cite{cru24}, GC masses from \cite{bau18} and, when not available, from \cite{har96} (2010 edition) V luminosities assuming $(M/L)_V=1.6$.    
   }
\label{fig:1}
\end{figure}

The function $\Phi(M)$ has been deduced from a suitable power-law interpolation of data in Fig.\ref{fig:1}, leading to

\begin{equation}
\label{eq:3}\Phi(M)=
\begin{cases}
aM^{\alpha},~ M_m\leq M\leq M_M,\\
0,~~~~ M<M_m ~or~ M>M_M,
\end{cases}
\end{equation}
    
where $M_m$ and $M_M$ are the lower and upper mass cutoffs (in solar masses), assumed as 
$M_m=1.06\times 10^4$M$_\odot$ and $M_M=3.55\times 10^6$M$_\odot$, respectively, 
and with $a=6.286\times 10^{-7}$ and $\alpha=0.428$, as coming from a logarithmic least square fit to the data (see Fig. \ref{fig:1}).
The Pearson correlation coefficient, $r$,  for the $N$ vs $M$ relation is $r=0.41$, meaning a sufficiently clear positive correlation, although with the large scatter shown in Fig.\ref{fig:1}.
As example, a relative perturbation $\Delta \alpha /\alpha$ of the exponent $\alpha$ around the value $\alpha=0.428$ in Eq.\ref{eq:3} leads to a relative variation of the number $N(<M)$,

\begin{equation}
\label{eq:4}
\dfrac{\Delta N(< M)}{N_0}= 1.159\times 10^{-10} M^{1.428}\left(3.288 \log M - 1 \right) \dfrac{\Delta \alpha}{\alpha},
\end{equation}

which means a propagation factor of $\Delta \alpha / \alpha$  below $5.4$  over the whole range $M_m\leq M \leq M_M$.

The integral over $M$ of $\phi(M)$ gives $N_{0} \simeq 1002.75$.  
A linear scaling implies that to get the observed number of RR Lyrae ($N_{RR}$) a number of $N_{GC} =(111/1002.75) N_{RR}$ globular clusters are needed.\par\noindent


\citet{ASCD17} evaluated the combined role of dynamical friction and tidal disruption on a set of GCs in a galaxy modeled as in Eq.\ref{eq:2} and, by a simple manipulation of their Eqs. 5, 6 and 7, $\mbar$ as a function of $\rbar,r,e,\gamma,\tau,M_{m},M_g,r_g$ is obtained as

\begin{equation}
\label{eq:5}
\mbar=\max\left(M_m,M_gA^\delta\left[\bar{x}^\beta-x^\beta\right]^\delta\right),
\end{equation}




where $A=A(e,\gamma,\tau)$, $\beta =1.76$, $\delta=3/2$, $x\equiv r/r_g$, $\bar{x}\equiv \bar{r}/r_g$.
The dimensionless function $A(e,\gamma,\tau)$ (where, as we said, $e$ is the GC orbital eccentricity and $\gamma$ the Galactic Dehnen's density profile slope) is the ratio to $\tau$ of the product of the function $t_0(M_g,r_g)$ and the $g(e,\gamma)$ as given by Eqs. 7 and 8 of \cite{ASCD17}, that is

\begin{equation}
\label{eq:6}
A\equiv \frac{t_0g}{\tau} =\frac{0.3}{\tau} \frac{\left(\frac{r_g}{1{\rm kpc}}\right)^{3/2}\left(\frac{10^{11}{\rm M}_\odot}{M_g}\right)^{1/2}}{(2-\gamma)\left[a_1\left(\frac{1}{(2-\gamma)^{a_2}}+a_3\right)(1-e) + e\right]},
\end{equation}

with $a_1=2.63\pm0.17$ and $a_3=0.9 \pm 0.1$.

\begin{table}
	\centering
	\caption{Parameters characterizing 8 of the models presented in sect. \ref{sec:3}
 In each model eccentricity is varied: $e=0, 0.1, 0.3, 0.5, 0.9$.}
 	\label{tab:1}
	\begin{tabular}{|ccccr|} 
		\hline
		mod. & $\gamma$ & $\tau$ (Gyr) & $r_g$ (kpc) & $x_R$\\
		\hline
		1 & 0.1 & 13.7 & 10 & 0.5\\
		2 & 0.1 & 13.7 & 10 & 1\\
            3 & 0.1 & 13.7 & 10 & 0.25\\
            4 & 0.1 & 13.7 &  2 & 1.25\\
            5 & 0.1 & 13.7 & 10 & 0.2\\
            6 & 0.1 & 13.7 &  2 & 1\\
            7 & 0.5 & 13.7 & 10 & 0.5\\
            8 & 0.5 & 13.7 & 10 & 1\\
            9 & 0.1 & 13.7 & 2 & 0.2\\
            10 & 1 & 13.7 & 10 & 0.2\\
            \hline
	\end{tabular}
    
\end{table}

            \begin{table}
	\centering
	\caption{Pairs of half-projected number and half-projected density radii (in kpc) at $\tau = 13.7$ Gyr for  $e=0.0, 0.5, 0.9$ (column pairs from left to right, separated by a double vertical line). \citet{Navetal21} give for the MW $R_h=0.99$ kpc and $R_{\Sigma h}=0.25$ kpc.}
	\label{tab:2}
	\begin{tabular}{|c||c|c||c|c||c|r|} 
	\hline
         mod. & $R_h$ & $R_{\Sigma h}$ & $R_h$ & $R_{\Sigma h}$ & $R_h$ & $R_{\Sigma h}$ \\
		\hline
		 1 & 2.68 & 3.69 & 2.60 & 3.51 & 2.57 & 3.36 \\
          2 & 5.81 & 8.36 & 5.70 & 8.16 & 5.50 & 7.68 \\
          3 & 1.28 & 1.66 & 1.28 & 1.65 & 1.27 & 1.64 \\
          4 & 6.43 & 8.46 & 6.43 & 8.46 & 6.37 & 8.24 \\
          5 & 1.02 & 1.32 & 1.02 & 1.32 & 1.02 & 1.31 \\
          6 & 1.02 & 1.33 & 1.02 & 1.32 & 1.02 & 1.32 \\
          7 & 2.65 & 3.64 & 2.59 & 3.46 & 2.26 & 3.34 \\
          8 & 5.81 & 8.36 & 5.65 & 8.06 & 5.50 & 7.68 \\
          9 & 0.204 & 0.263 & 0.204 & 0.263 & 0.204 & 0.263\\ 
          10 & 1.02 & 1.32 & 1.02 & 1.32 & 1.02 & 1.31\\   
	\hline
	\end{tabular}
    \end{table} 


\begin{table}
	\centering
	\caption{The values of $N(<35 \,\textrm{pc})/N_0$ and the corresponding estimates of $N_0$ to give the estimated value of $N(<35 \,\textrm{pc})= 1562$ of the MW, for the models of Tab. \ref{tab:1} with $e=0.5$.}
	\label{tab:3}
	\begin{tabular}{ccc} 
		\hline
		mod. & $N(<35 \,\text{pc})/N_0$ & $N_0/10^6$ \\
		\hline
		1 & $1.49\times 10^{-4}$ & 10.5 \\
		2 & $3,31\times 10^{-5}$ & 47.2 \\
            3 & $6.59\times 10^{-4}$ & 2.37 \\
            4 & $6.52\times 10^{-4}$ & 2.40 \\
            5 & $1.03\times 10^{-3}$ & 1.51 \\
            6 & $7.28\times 10^{-4}$ & 2.14 \\
            7 & $1.53\times 10^{-4}$ & 10.2 \\
            8 & $2.77\times 10^{-5}$ & 56.4 \\
            9 & $2.01\times 10^{-3}$ & 0.78 \\
            10 & $1.04\times 10^{-2}$ & 0.15 \\
            \hline
	\end{tabular}
    
\end{table}

Putting together Eqs. \ref{eq:1} ,\ref{eq:2}, \ref{eq:3}, \ref{eq:5},\ref{eq:6} eventually leads to

\begin{equation}
\label{eq:7}
N(<x;\tau) = N_{0}\Biggl[{\left(\dfrac{x}{x_R}\right)^3+ \dfrac{3 a}{\alpha+1}\dfrac{1}{x_R^3N_{0}}\int\limits_{x}^{x_R}{\bar x}^2\left(M_M^{\alpha+1}-\mbar^{\alpha+1}\right)d \bar x
\Biggr]},
\end{equation}

so that the spatial number density distribution of the RR Lyrae is obtained by the simple relation

\begin{equation}
\label{eq:8}
\rho(x;\tau)=
\frac{1}{4\pi r_g^3 x^2} \frac{dN}{dx},
\end{equation}

which can be calculated from Eq. \ref{eq:7}.

The corresponding surface (projected) numerical density is obtained via

\begin{equation}
\label{eq:9}
\Sigma(\tilde s;\tau)= 2r_g\int\limits_{\bar s}^{x_R}\frac{\rho(x)x}{\sqrt{x^2-\tilde s^2}}dx.
\end{equation}

where $\tilde s=s/r_g$ is the dimensionless projected radial distance to the center. For numerical convenience, the improper integral above is transformed into a proper one by an integration by parts, letting $u(x)=\rho(x)$ and $dv(x)= \left( x^2-\tilde s^2\right)^{-1/2}x dx$, leading to

\begin{equation}
\label{eq:10}
\Sigma(\tilde s;\tau)= -2r_g\int\limits_{\tilde s}^{x_R}\sqrt{x^2-\tilde s^2}\, \frac{d\rho}{dx} dx.
\end{equation}



\section{Results}
\label{sec:3}
Here I present a general discussion of the models and, after, some considerations on the observational-model comparison.
\subsection{Discussion of the models}
\label{sec:3.1}
With the method and formalism described above we can estimate (from Eqs. \ref{eq:3}-\ref{eq:10}) the expected distribution of RR Lyrae around the galactic center, computing $N(<s;\tau)$ (Eq. \ref{eq:7}) and the spatial (Eq. \ref{eq:8}) and projected densities (Eqs. \ref{eq:9} and \ref{eq:10}).\\
We have produced some models for various choices of the main parameters, namely $\gamma$, $r_g$, and $x_R=R/r_g$, varying the eccentricity $0\leq e \leq 0.9$, and over the age interval $0 \leq \tau \, \mathrm{(Gyr)} \leq 13.7$. The characteristics of the $10$ models studied (each referred to 5 different values of $e$) are reported in Table \ref{tab:1}.
For the sake of a better display of results, I chose to insert in the Appendix \ref{app:B} all the figures useful to understand the role of the various parameters and below discussed.

\begin{figure}
	\includegraphics[width=\columnwidth]{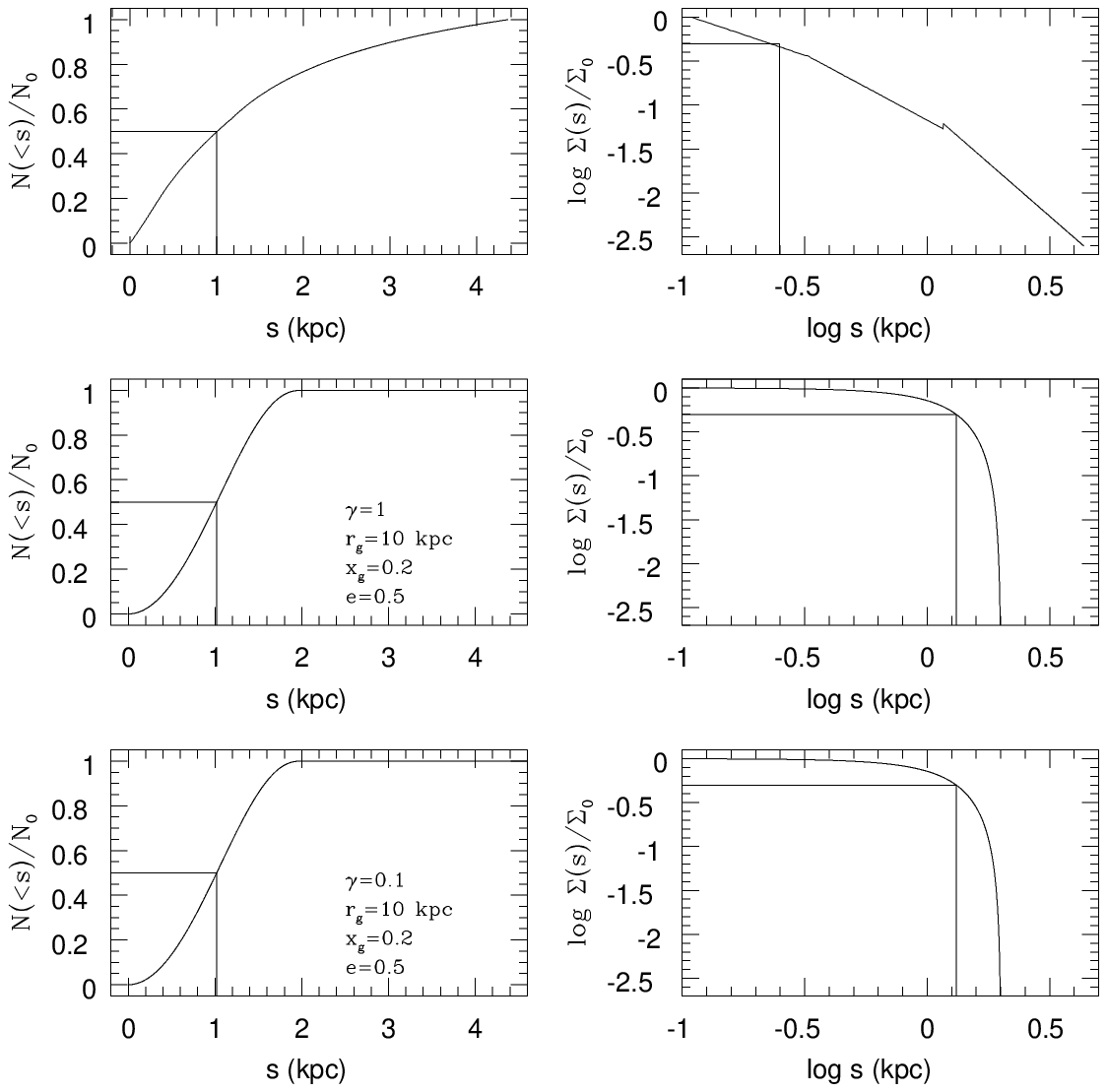}
        \caption{Comparison among the RRab spatial and surface distribution from \citet{Navetal21} (left and right top panels) and those corresponding to models with 
        the parameter values given in the left panels. The intersection of horizontal and vertical segments refers to half-number radius.}
    \label{fig:2}
\end{figure}

In Figs. \ref{fig:B2}-\ref{fig:B6} we report the profiles of $N(<\tilde s)/N_0$ at the times $\tau \, \mathrm{(Gyr)} =0.1,\, 1,\, 5,\, 13.7$,  assuming, in Eq. \ref{eq:6} $M_g=2\times 10^{11}$ M$_\odot$, three values of $\gamma$ ($0.1$, $0.5$ and $1$), two values for the parameter $r_g$ ($2$ and $10$ kpc), three values for $x_R=R/r_g$ ($0.2, 0.5, 1$), and four values for $e$, namely $0,\, 0.3,\, 0.5,\, 0.9$, with the color code indicated in the Fig. \ref{fig:B2} caption.\\ 
Figures \ref{fig:B2}-\ref{fig:B6} show a very slight dependence of $N(<\tilde s)$ upon the GC orbital eccentricity; as expected, larger $e$  (more radially pointed orbits) implies a more compact evolved radial distribution, but this is limited to a variation which is appreciable only over various Gyrs of evolution. The maximum fractional increment of the half-number radius is $9.7$\% going from $e=0$ to $e=0.9$ (case $\gamma=0.1$, $r_g=10$ kpc, $x_R=1$, i.e. mod. 2 at $\tau=13.7$ Gyr).

The dependence of $N(<\tilde s)/N_0$ upon $\gamma$ in the three cases for $\gamma$ studied is marginal, too, as visible by a comparison between mod. 5  and mod. 10 (Fig. \ref{fig:B4}), and confirmed by values in Table \ref{tab:2} which shows a variation of less than $1\%$ for both $R_h$ and $R_{\Sigma h}$ between model 5 and model 8. This minor dependence on $\gamma$ convinced us to refer to $\gamma =0.1$ as fiducial value for all further considerations.

The dependence on $r_g$ and $x_R$ is almost linear, as expected. At the same time, the minor dependence on $e$ leads us to refer to the intermediate value $e=0.5$ as reference value for $e$. 
Figure \ref{fig:B7} shows the radial space density profile (upper four panels) and the projected density (lower four panels) at time $\tau =13.7$ Gyr assuming $r_g=10$ kpc and the values of $\gamma$ and $x_R$ as labeled.
It is notable the off-center peaks of $\rho(r)$, which are characteristics of the dynamical friction evolution. Projection, giving $\Sigma(s)$ from $\rho(r)$, cancels this feature and the surface density shows a monothonically decreasing behaviour with an inner plateau. Differences between $\gamma =0.5$ and $\gamma=1$ are negligible, while increasing $x_R$ moves rightward the spatial density peak and enlarges the surface density plateau.
Figure \ref{fig:B8} sketches the time evolution of the half-number projected radius for the two values of $\gamma$ ($\gamma =0.1$ in the upper 4 panels and $\gamma=0.5$ in the lower 4) and two values of both $r_g$ and $x_R$, as labeled.


\subsection{Observation-model comparison}

Figure \ref{fig:2} gives the observed RRab distributions from \cite{Navetal21} in cumulative number and projected density (upper two panels).  These data provide what is at present  the  RRab distribution most extended in radius of the MW,  from $\simeq 100$ pc from the center to  $\simeq 4.8$ kpc  in the halo.
Figure \ref{fig:2} gives the distributions from model 10 (intermediate panels) and model 5 (lower panels) both for $e=0.5$. I chose these two models because they fits  reasonably well the half-number and half-projected density radii of the Galactic RRab (see Table \ref{tab:2}.
The inner behaviour of the observational RRab distribution is different from our theoretical one because of the limited inner resolution of observational data whose innermost trend is just an extrapolation inward with a $-1$ slope. This explains also the flatter $\Sigma(s)$ distribution of theoretical models respect to the observational fitted profile (right panels in Fig. \ref{fig:2}).

The values of the ratio $R_h/R_{\Sigma h}$ of the models studied here (Table \ref{tab:2}) always correspond to $R_h<R_{\Sigma h}$ while the MW data give $R_h\simeq 4R_{\Sigma h}$. This is indeed explained by the above-mentioned flatter radial distribution of the models.
Actually, as described in the Appendix \ref{app:A} and sketched in Fig. \ref{fig:A1}, the ratio $R_h/R_{\Sigma h}$
when evaluated for decreasing power law distributions ($\rho \propto r^\alpha, \alpha <0$) and taking also into account the, unavoidable, lower radial cut-off is (for any reasonable ratio of the upper to lower radial cut-off $s_M$ and $s_m$) above $1$ for steeper power laws including the $\alpha \simeq -1$ obtained in \cite{Navetal21} for the MW data, while it is below $1$ for flatter (\textit{cored}) distributions.
   Note that the halving value for $\Sigma(s)$ obtained in \cite{Navetal21} is evaluated with respect to a `central' value which is that at $s_m \simeq1^\circ$ (= 0.15 kpc). 


 Finally, I reported in Table \ref{tab:3} the values of the fraction of RR Lyrae enclosed in a 35 pc radius from the Galactic center for the 8 models of Table \ref{tab:1} with $e=0.5$, and the total $N_0$ needed to reproduce the actual estimated number of RR Lyrae in this inner region (which is $1562$, corresponding to a number density $\simeq 0.02$ pc$^{-3}$ \citep{Navetal21}), which ranges from $1.5\times 10^5$ to $5.64 \times 10^7$. Adopting the linear scaling of RR Lyrae in a GC as that of Sect. \ref{sec:2}, this implies a primeval GC population $1.66\times 10^4 \leq N_{GC} \leq 6.24 \times 10^6$. Given the present number of  observed  GCs in the Milky Way, $N=158$ \citep{Bau19} this means a survival percentage fraction $0.003 \leq f (\%) \leq 0.95$, that is very small.
 At this regard it is relevant noting that at these distances from the center ($\leq 35$ pc) star formation may still be ongoing in the nuclear stellar disc. although it is likely composed mainly by old stellar populations \citep{sch23}.

\section{Conclusions}\label{sec4}

In this paper it has been discussed the possibility that the observed distribution of type ab RR Lyrae stars in the Galactic bulge come from dissolved globular star clusters, whose system has dynamically evolved mainly due to dynamical friction acting on their orbital paths so to release part of its stars to the environment.

The observational data used here are coming mainly from the \cite{Navetal21} paper while models rely mainly on re-elaboration of \cite{ASCD17} work. The comparison among the projected cumulative number and density distributions, $N(<s)$ and $\Sigma(s)$ respectively, indicates a flatter radial projected density distribution of the theoretical models respect to the observational data. This is partly due to the limited inward extension of data and their subsequent inner extrapolation by a power law (of slope $\simeq -1$), obviously divergent to the center). More robust is the comparison between the cumulative $N(<s)$ distributions which, in the models, show halving values, $R_h$, having a weak dependence on both the $\gamma$ exponent in the Dehnen's model profile adopted for the Galaxy bulge/halo and on the orbital eccentricities, while the dominant dependence is on the $r_g$ (the Dehnen's model scale length) and on $R$ (initial maximum radial extension of the GC system) scale lengths.\\

The main results of this paper are:
\begin{itemize}
\item 
a more stringent quantitative comparison is at present difficult to do because of the lack of observations in the inner Galactic region, where extrapolations of data surely fail;
\item
at the light of previous point, the observed distribution of RR Lyrae stars in the bulge shows qualitative compatibility with being heritage of such stars as belonging to parent globular clusters whose orbital evolution in the Galaxy has been primarily dictated by dynamical friction;
\item 
to justify by their origin from GCs, the present estimates of the number of RR Lyrae present in the central region of the MW require either the number of primordial GCs was very large or their individual RR Lyrae content much greater than what presently observed.
\end{itemize}

\bmhead{Acknowledgements}
I thank Dante Minniti for useful discussions on the topic of this work.



\noindent



\bibliography{biblio_apss}

\vfill\eject
\begin{appendices}

\section{The $R_h-R_{\Sigma h}$ ratio}
\label{app:A}



Given a truncated power law for the surface density $\Sigma(s)$

\begin{equation}
\label{eq:A1}\Sigma(s)=
\begin{cases}
as_m^\alpha, ~ 0\leq s\leq s_m,\\
as^{\alpha},~ s_m\leq s\leq s_M,\\
0,~~~~ s > s_M,
\end{cases}
\end{equation}

with $\alpha <0$, the central half value distance $R_{\Sigma h}$ is obtained as 
$R_{\Sigma h}=\sqrt[\alpha]{1/2} s_m$.
The above surface density distribution implies 

\begin{equation}
    N(<s)=
    \begin{cases}
    2\pi a \int\limits_{s_m}^{s} s^{\alpha+1} ds=
    2\pi a \,\dfrac{s^{\alpha+2}-s_m^{\alpha+2}}{\alpha+2}, \,\, & s_m \leq s\leq s_M,\, \alpha\neq -2,\\
    2\pi a \int\limits_{s_m}^{s} \dfrac{ds}{s}=2\pi a \ln{(s/s_m)}, \,\,\,\,\, & s_m \leq s\leq s_M,\, \alpha= -2,
    \end{cases}
\end{equation}

corresponding to a total number

\begin{equation}
    N(\leq s_M)=
    \begin{cases}
     2\pi a \, \dfrac{s_M^{\alpha+2}-s_m^{\alpha+2}}{\alpha+2}, & \alpha\neq -2,\\
    2\pi a \ln{(s_M/s_m)}, & \alpha= -2.
    \end{cases}
\end{equation}

\begin{figure}
	\includegraphics[scale=0.6]{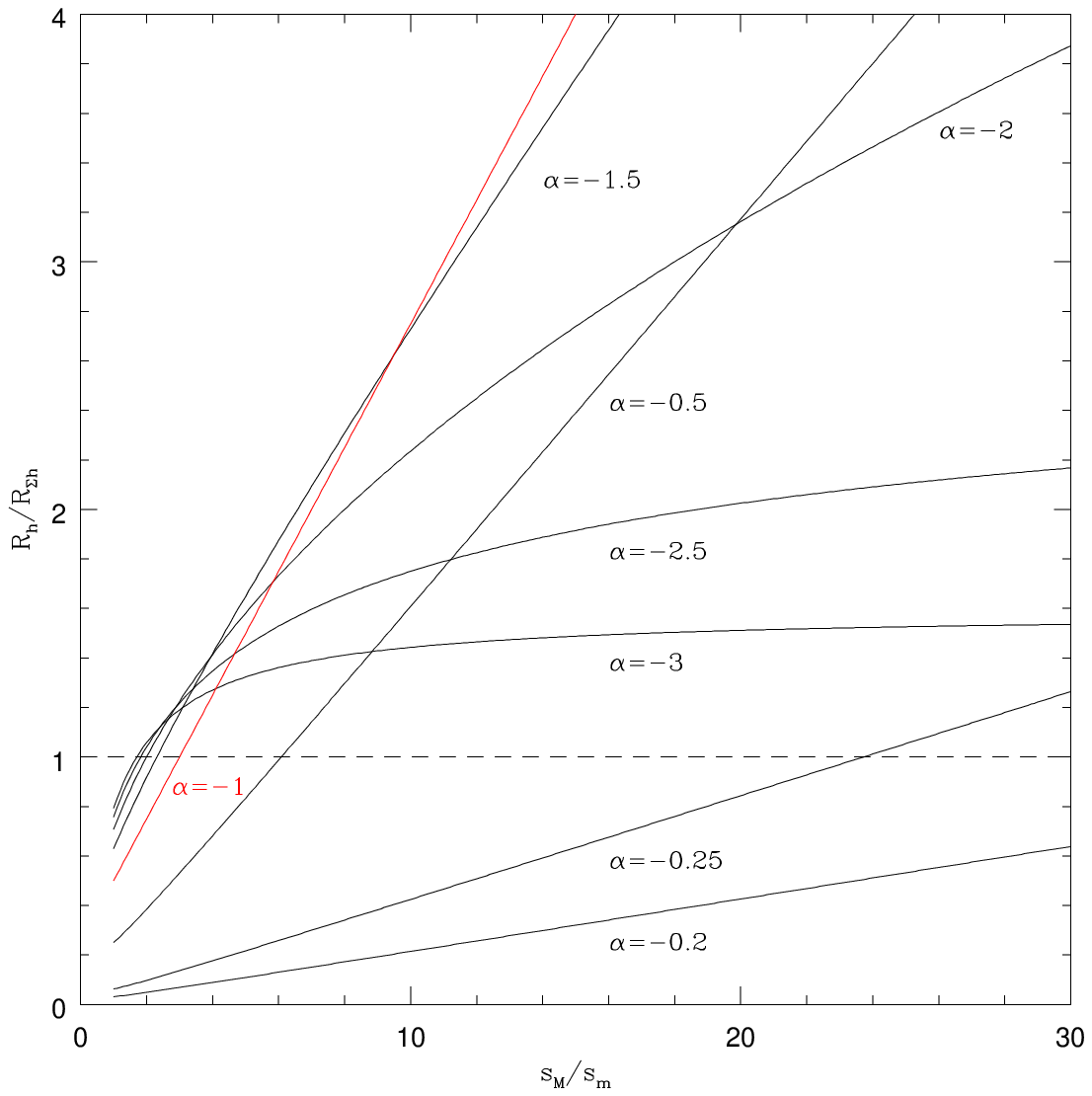}
        \caption{Ratio of the half-number to half-surface density radii for various exponent power laws (Eq. \ref{eq:A1}).
        Each curve is labeled by its $\alpha$ value.
    The black dashed horizontal line is $R_h=R_{\Sigma h}$.}
    \label{fig:A1}
\end{figure}

 Consequently the half-number distance $R_h$ is obtained as

 \begin{equation}
 R_h=
 \begin{cases}
 \sqrt[\alpha+2]{(s_M^{\alpha+2}+s_m^{\alpha+2})/2}, & \alpha\neq -2,\\
 \sqrt{s_Ms_m}, & \alpha=-2,
 \end{cases}
\end{equation}



and so the ratio of the half-number to the half-surface density distance to the center is

\begin{equation}
\frac{R_h}{R_{\Sigma h}}=
\begin{cases}
\left\{4^{\frac{1}{\alpha}}\left[ \left(\frac{s_M}{s_m}\right)^{\alpha+2}+1 \right]\right\}^{\frac{1}{\alpha+2}}, & \alpha\neq -2 \\
\sqrt{\frac{1}{2}\frac{s_M}{s_m}}, & \alpha=-2.
\end{cases}
\end{equation}

Figure \ref{fig:A1} shows that in the \lq steep\rq \, power law regime $\alpha \leq -0.5$ (which includes the inner $\alpha=-1$ MW slope) 
$R_h> R_{\Sigma h}$ whenever $s_M/s_m \geq 6.08$, the crossing point ranging from $s_M/s_m \simeq 1.69$ for $\alpha=-3$ to $s_M/s_m \simeq 6.08$ for $\alpha =-0.5$. Shallower, almost cored, radial distributions on the contrary correspond to
$R_h <R_{\Sigma h}$.

\vfill\eject

\section{Plots referred to in sect. \ref{sec:3.1}}

\label{app:B}
\begin{figure}
\includegraphics[scale=0.6]{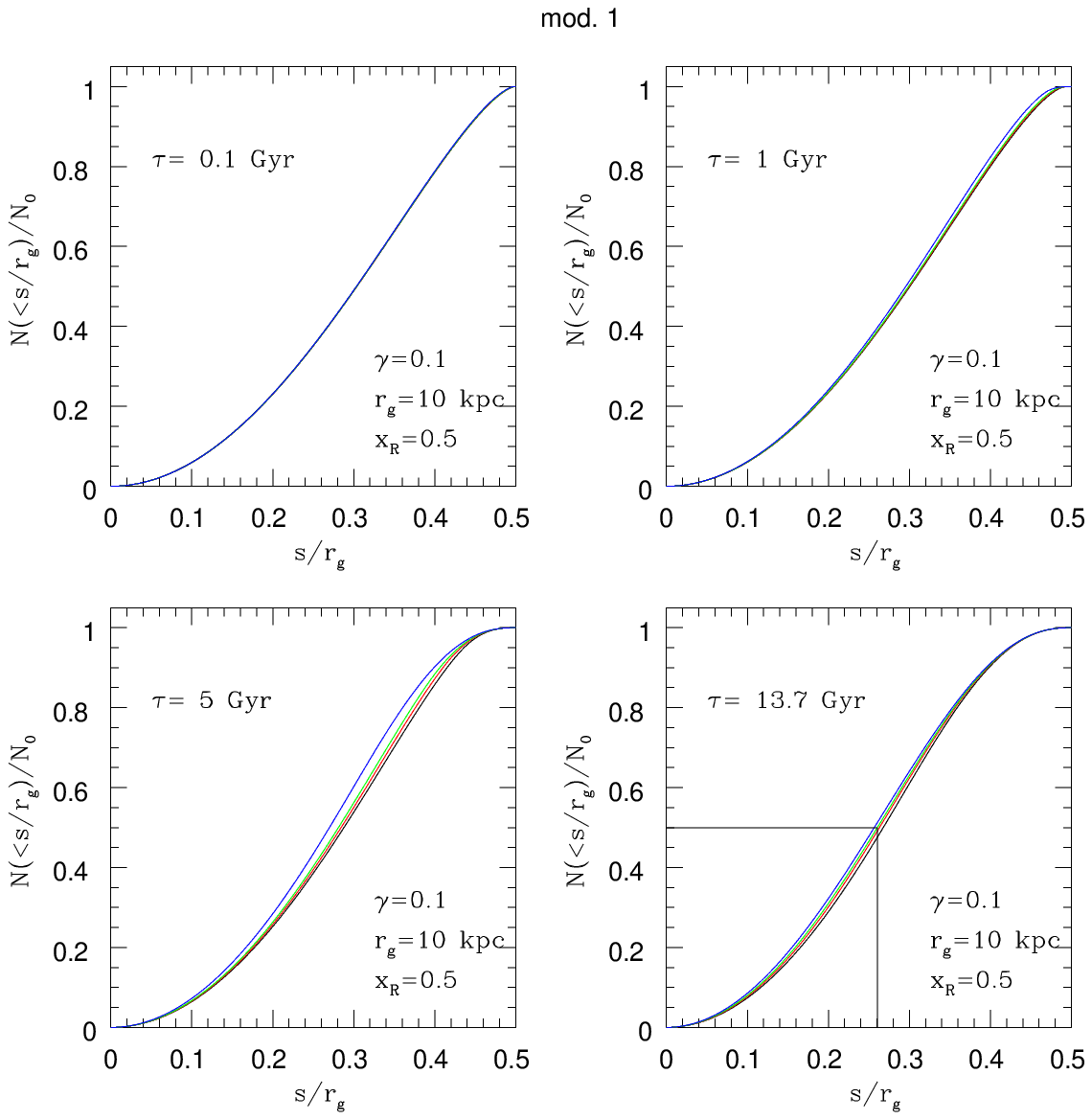}
\includegraphics[scale=0.6]{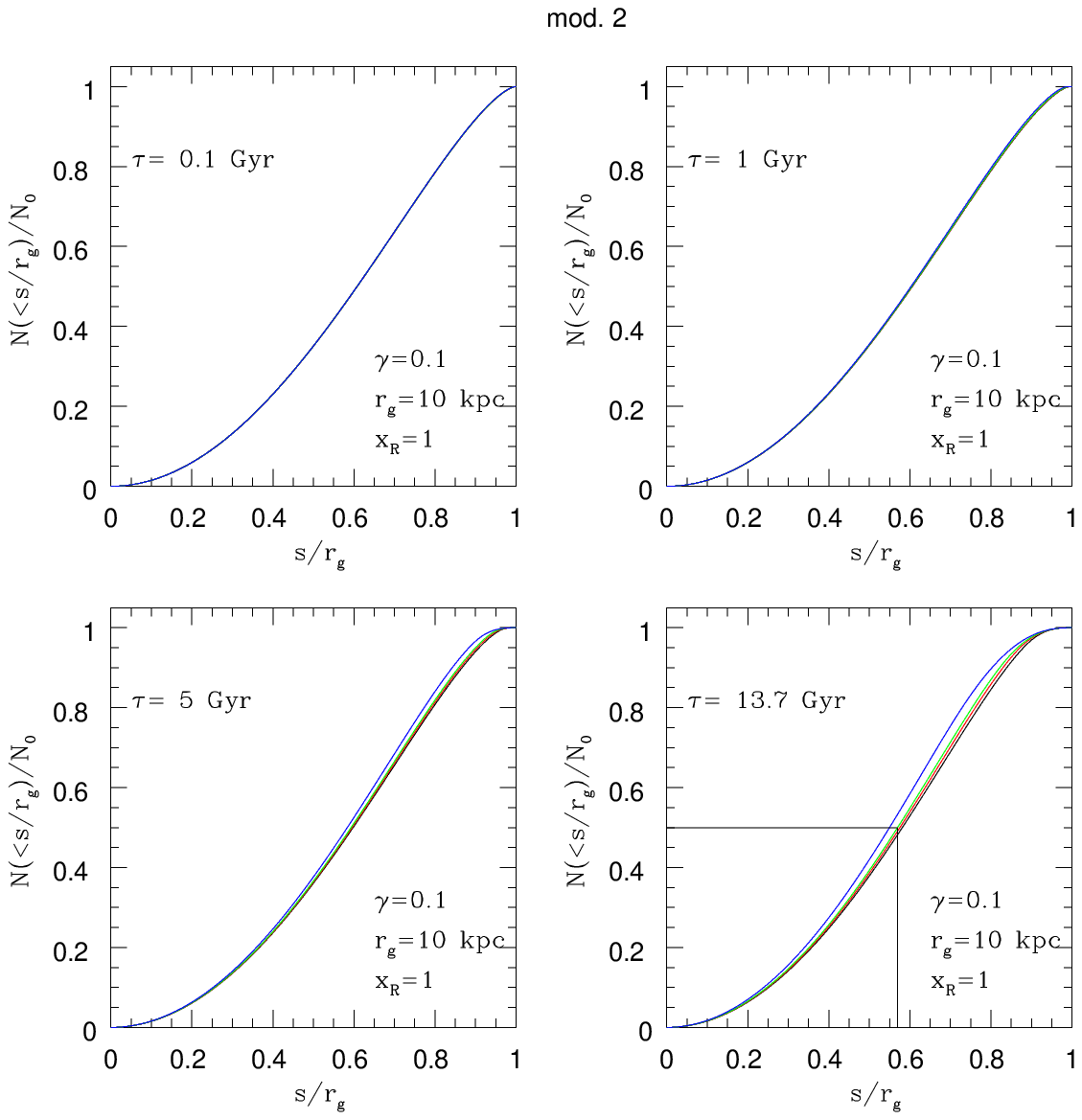}
    \caption{Radial profiles of $N(<s/r_g)/N_0$ at various times (as labeled) for 4 different orbital eccentricities: $e=0$ (black), $e=0.3$ (red), $e=0.5$ (green), $e=0.9$ (blue). The value of $\gamma$ is set to $0.1$ and $r_g=10$ while $x_g=0.5$ (upper 4 panels, model 1) and $x_g=1$ (lower 4 panels, model 2).
    The horizontal and vertical segments at $\tau =13.7$ Gyr refer to half-number radii for $e=0.5$. 
}
   \label{fig:B2}
\end{figure}

\begin{figure}
\includegraphics[scale=0.6]{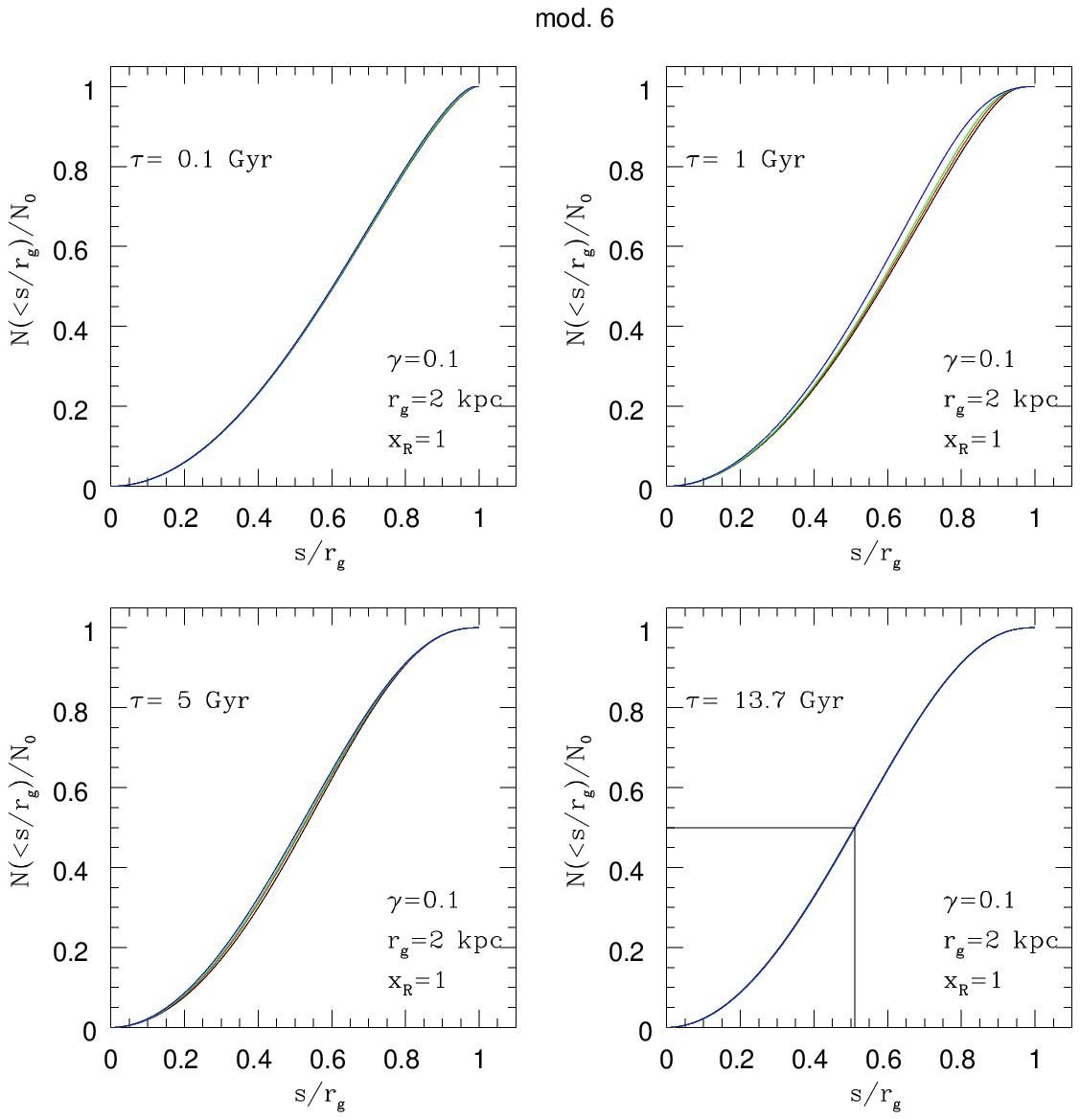}
\includegraphics[scale=0.6]{figures/number_g01_xR1_rg10.eps}
    \caption{As in Fig.\ref{fig:B2} but fixing $\gamma=0.1$ and $x_g=1$ and letting $r_g=2$ kpc (upper panels, model 6) and $r_g=10$ kpc (lower panels, model 2). 
}
    \label{fig:B3}
\end{figure}

\begin{figure}
\includegraphics[scale=0.6]{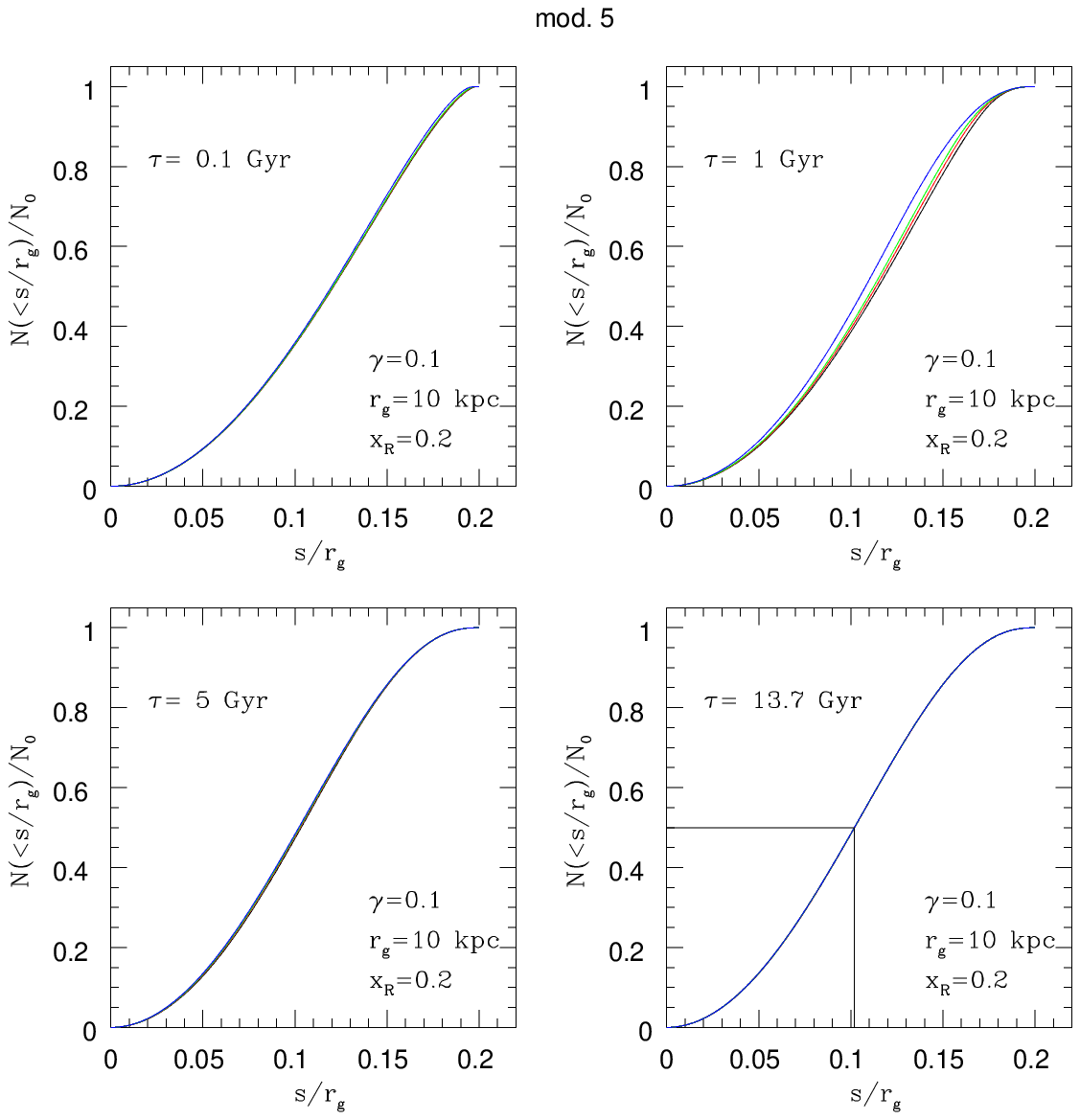}
\includegraphics[scale=0.6]{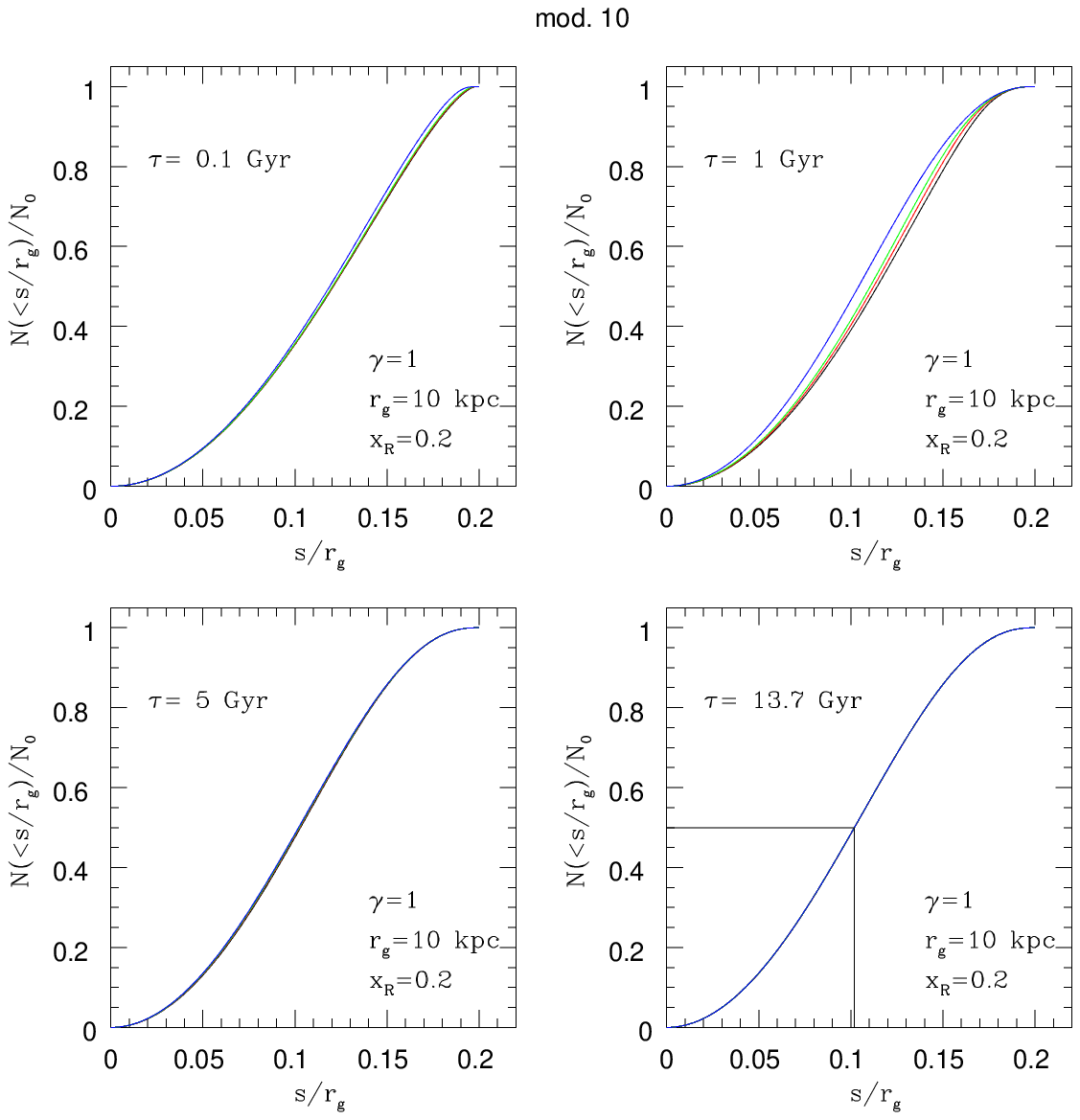}
    \caption{As in Fig.\ref{fig:B2} but fixing $r_g=10$ kpc and $x_g=0.2$ and letting $\gamma=0.1$ (upper panels, model 5) and $\gamma =1$ (lower panels, model 10). 
    }
   \label{fig:B4}
\end{figure}

\begin{figure}
	\includegraphics[scale=0.6]{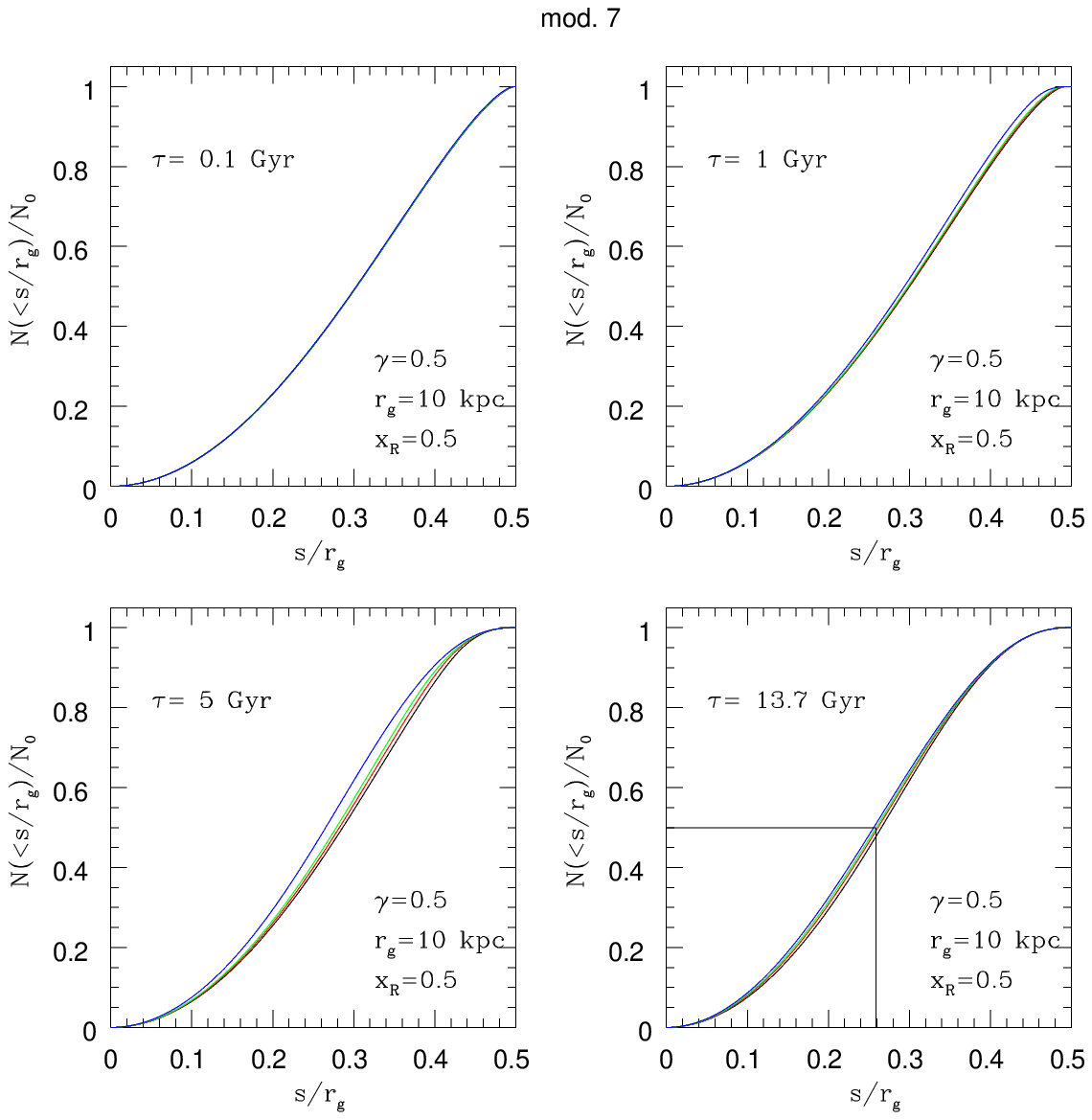}
        \includegraphics[scale=0.6]{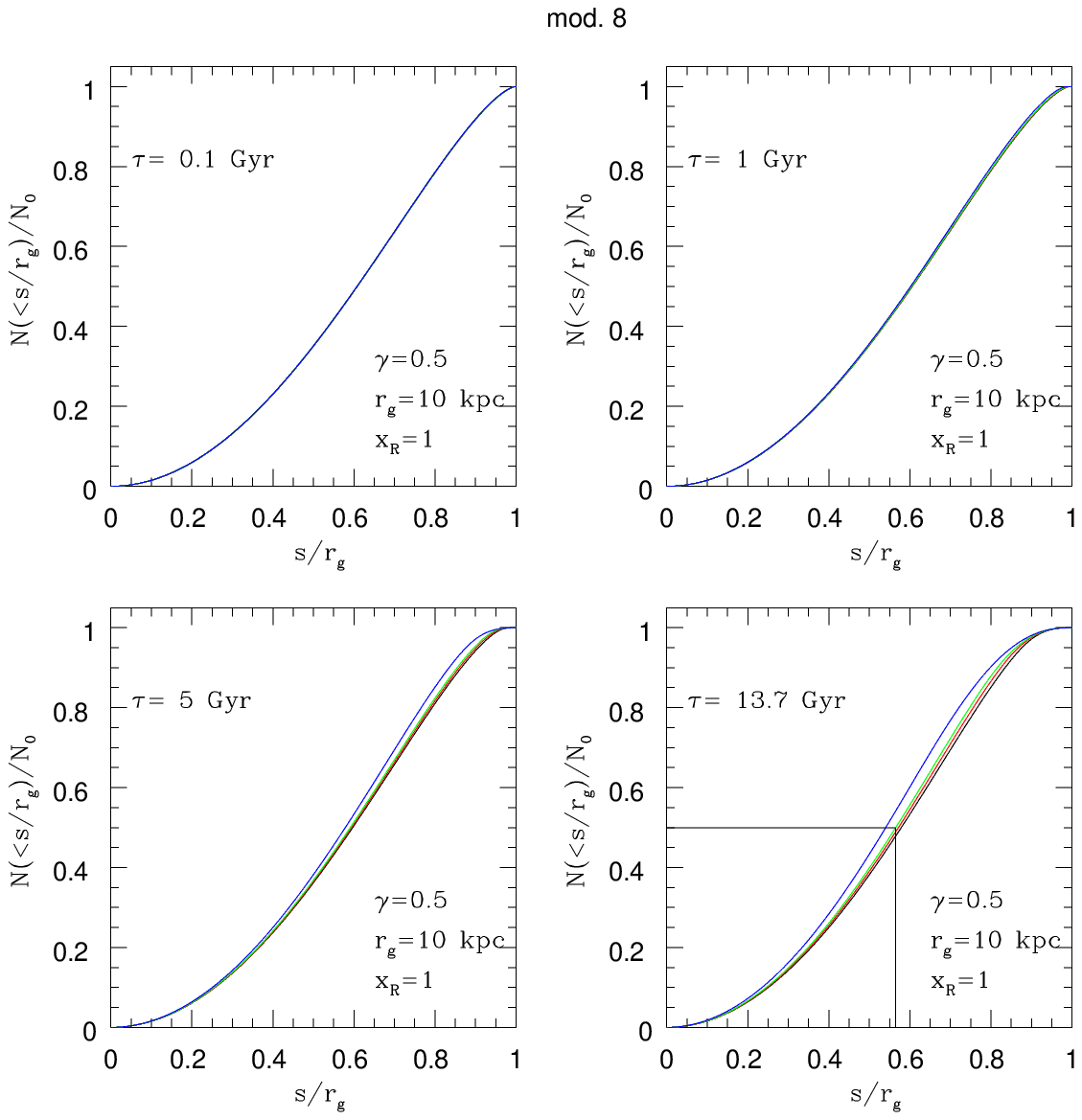}
    \caption{As in Fig. \ref{fig:B2} but fixing $\gamma=0.5$ and $r_g=10$ kpc and letting $x_R=0.5$ (upper panels, model 7) and $x_R=1$ (lower panels, model 8). 
    }
   \label{fig:B5}
\end{figure}

\begin{figure}
	\includegraphics[scale=0.6]{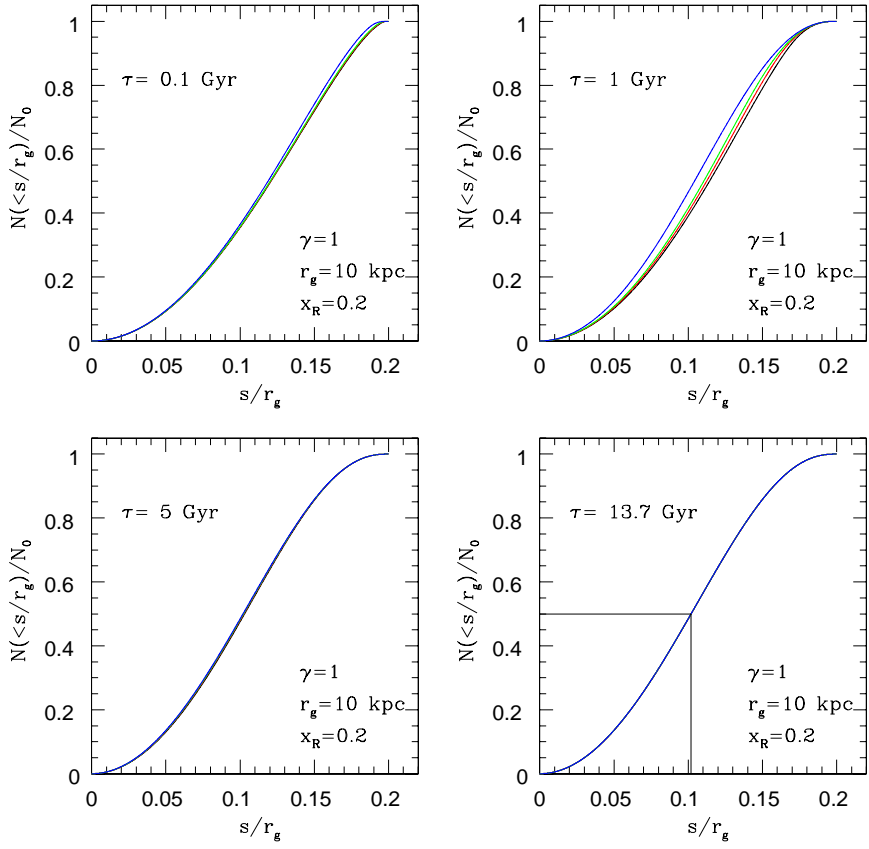}
        \includegraphics[scale=0.6]{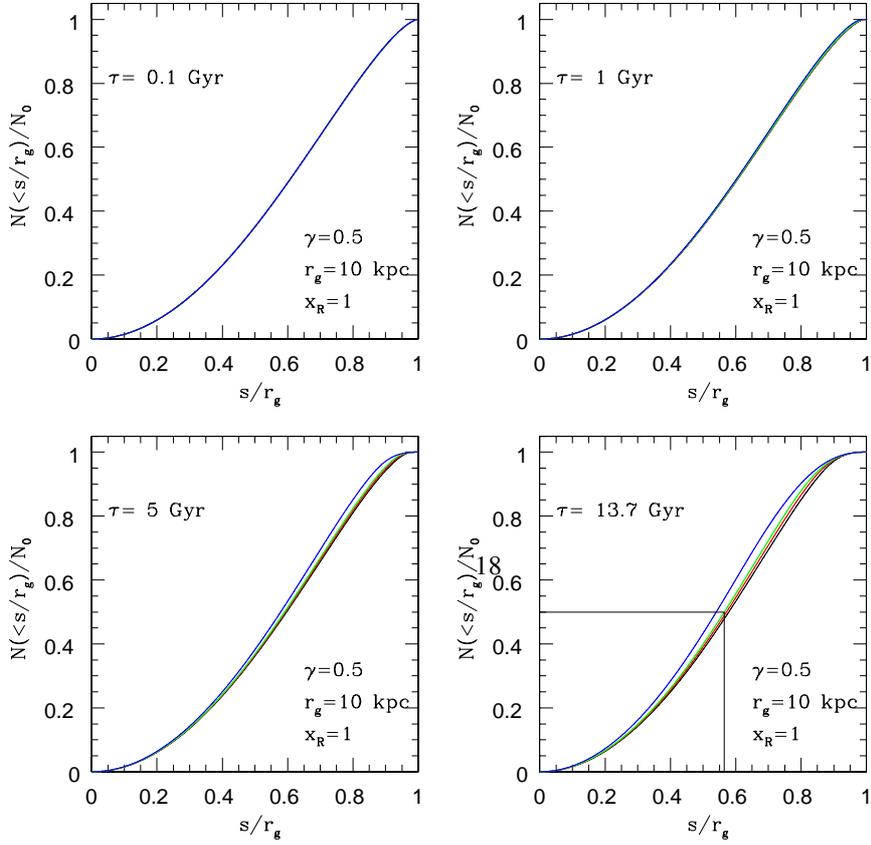}
    \caption{As in Fig. \ref{fig:B2}, for the cases of model 10 and model 8. 
    }
   \label{fig:B6}
\end{figure}

\begin{figure}
	\includegraphics[scale=0.6]{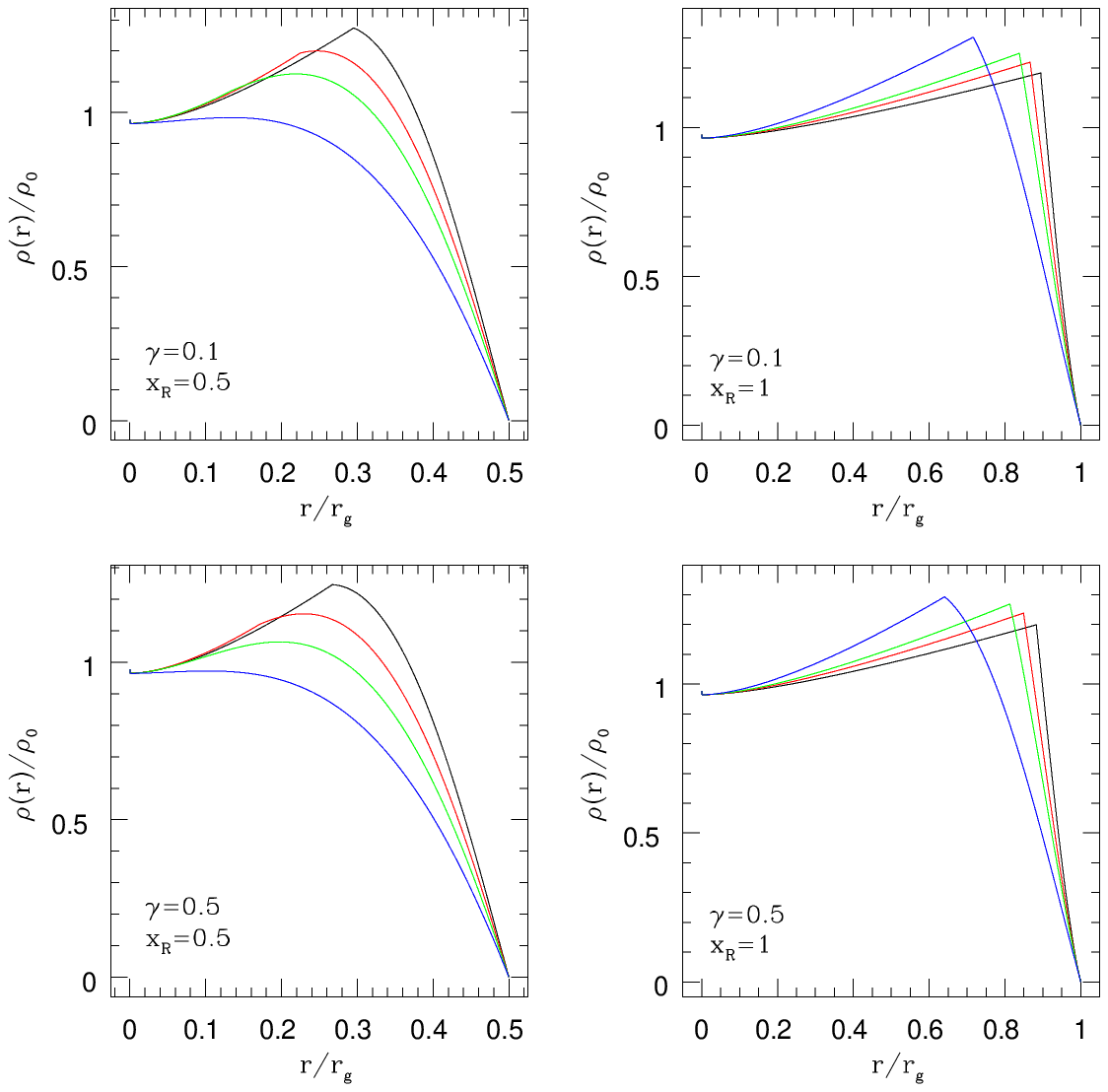}
        \includegraphics[scale=0.6]{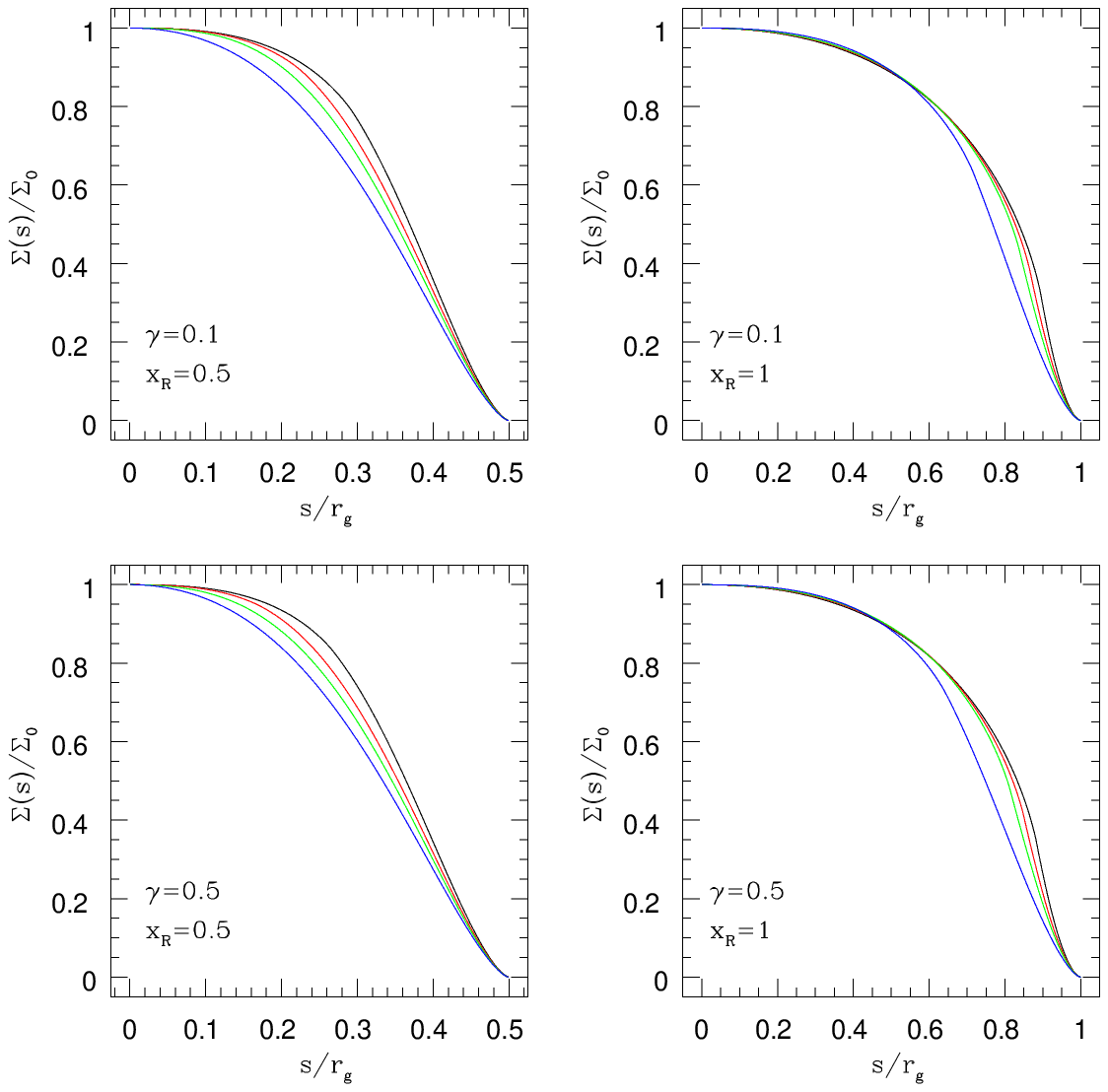}
    \caption{The upper 4 panels give the volume density radial profiles at present age ($\tau = 13.7$ Gyr) assuming $r_g=10$ kpc, for $\gamma =0.1$ and $\gamma=0.5$ and $x_R=0.5$ and $x_R=1$. 
    The lower 4 panels refer to the corresponding projected density. The eccentricity color coding is the same of previous figures.}
  \label{fig:B7}
\end{figure}

\begin{figure}
	\includegraphics[scale=0.6]{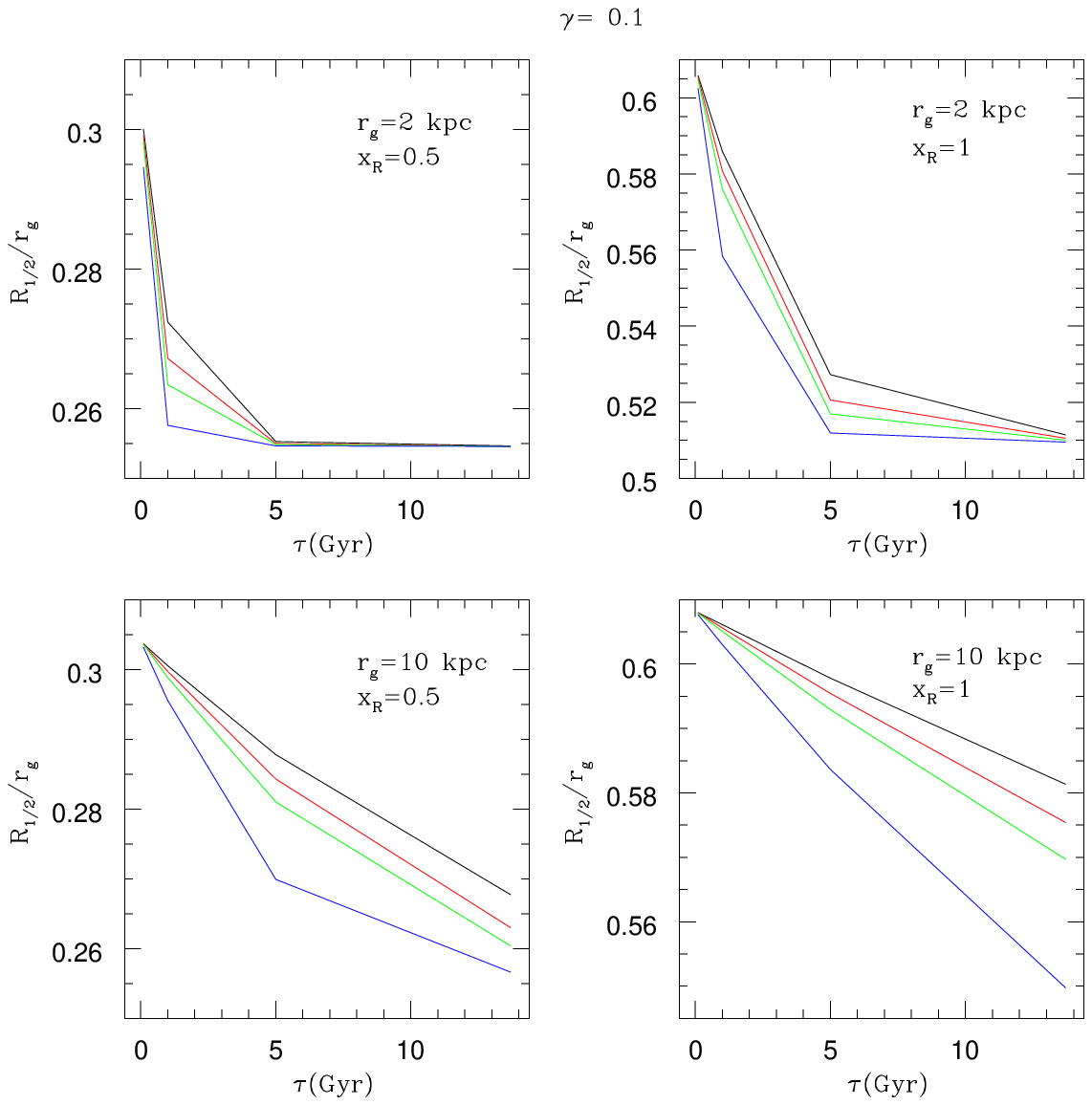}
        \includegraphics[scale=0.6]{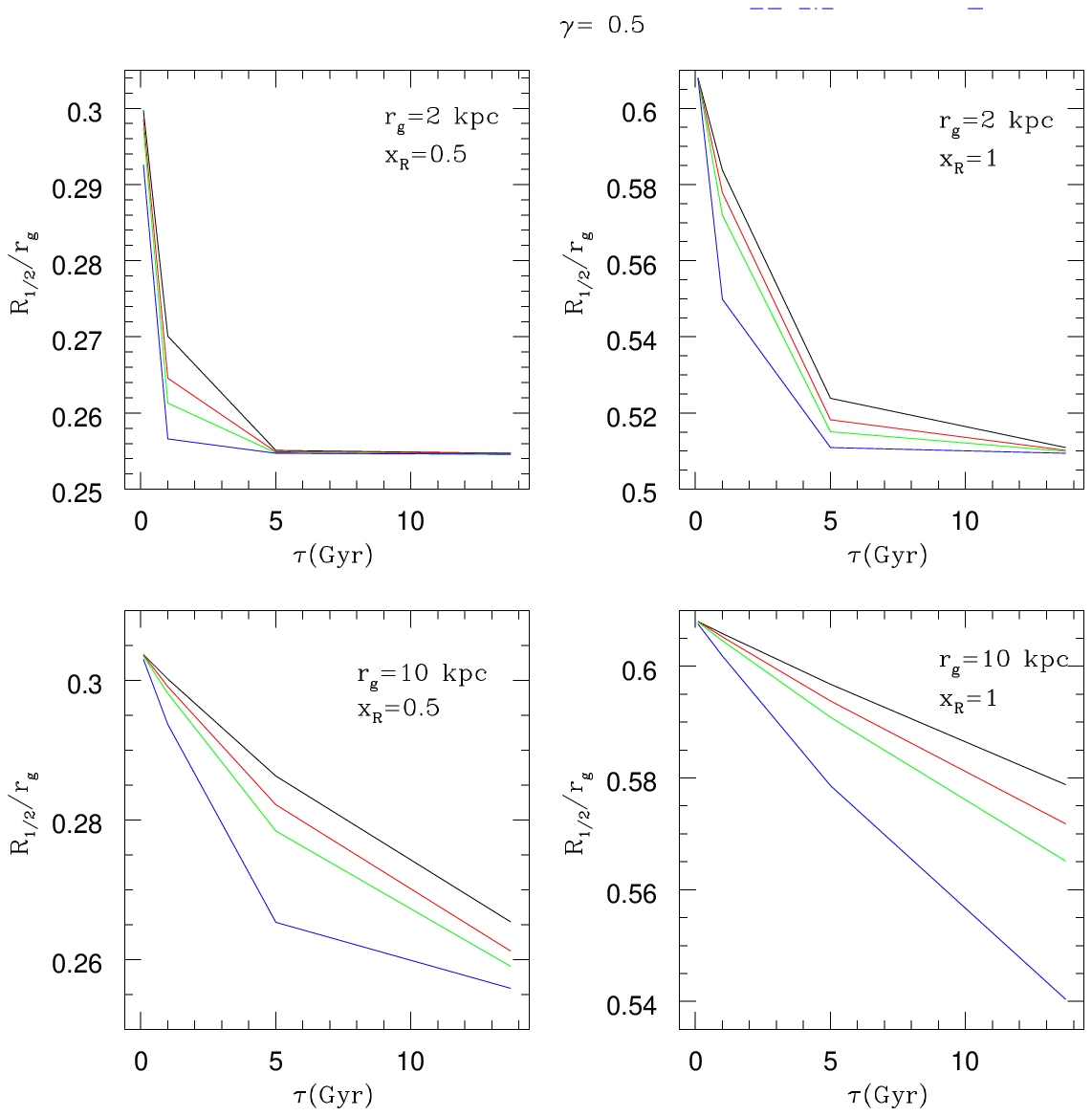}
    \caption{Time evolution of the projected half-number radius for $\gamma =0.1$ and $\gamma =0.5$ (set of 4 upper and lower panels, respectively). The various choices for $r_g$ and $x_g$ are labeled. The eccentricity color coding is the same of previous figures.}
    \label{fig:B8}
\end{figure}

\begin{figure}
	\includegraphics[scale=0.6]{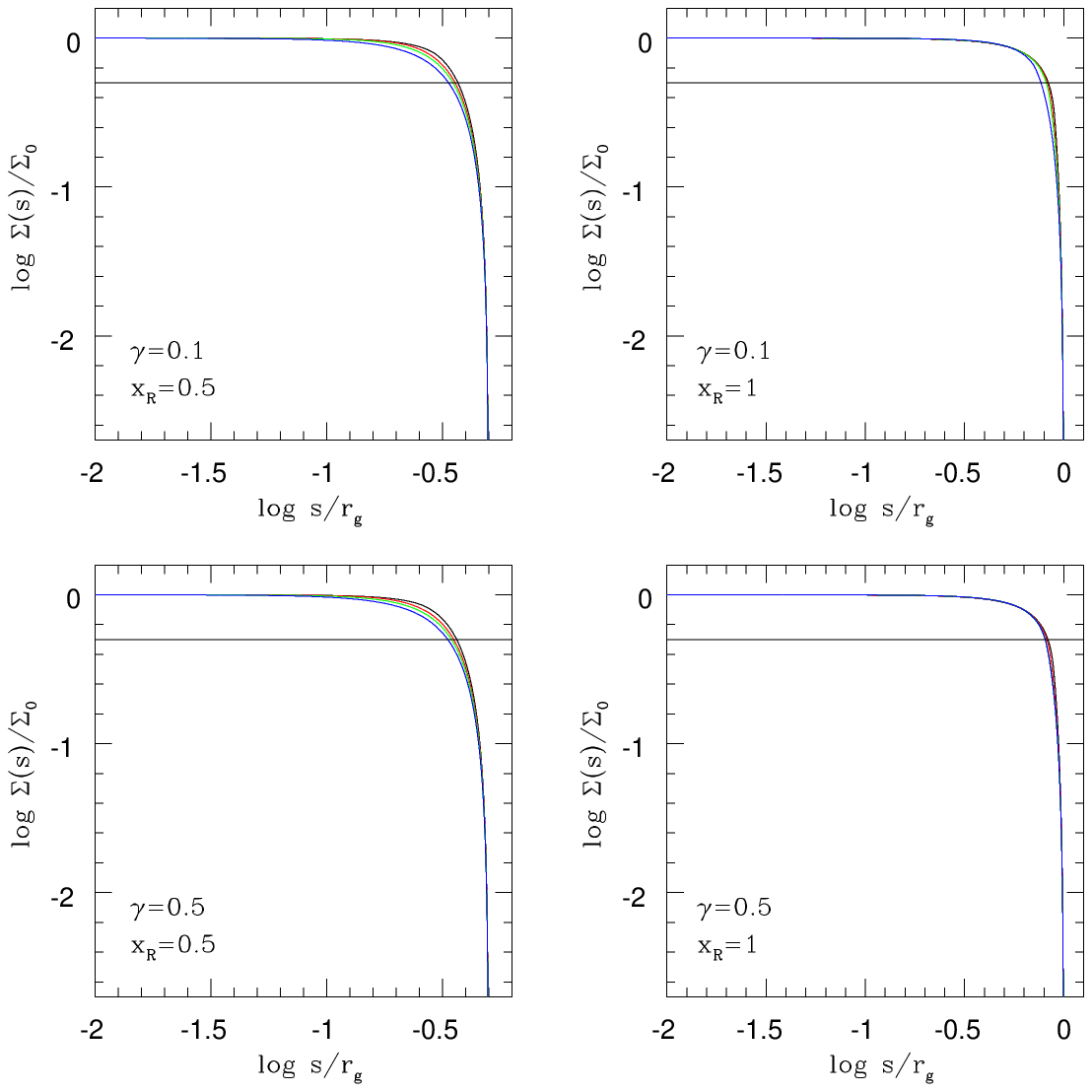}
       
    \caption{Surface density in bi-logarithmic plot of the same 4 cases of lower 4 panels in Fig. \ref{fig:B6}. The straight line is the central-halving value.}
    \label{fig:B9}
\end{figure}


\end{appendices}





\end{document}